\documentclass{ws-ijmpd}
\usepackage[super,compress]{cite}
\usepackage{graphicx}
\usepackage{amssymb}
\usepackage{hyperref}
\usepackage{url}
\usepackage{xcolor}
\usepackage{txfonts}

\begin{document}

\catchline{}{}{}{}{}
%
%%%%%%%%%%%%%%%%%%%%%%%%%%%%%%%%%%%%%%%%%%%%%%%%%%%%%%%%%%%%%%%%%%%%

\title{Expanding Microscopic Black Holes}

\author{Samuel Kováčik}

\address{Faculty of Mathematics, Physics and Informatics, Comenius University in Bratislava,\\
Mlynská dolina F1, Bratislava, 842 48, Slovakia\\
samuel.kovacik@fmph.uniba.sk}

\address{Department of Theoretical Physics and Astrophysics, Faculty of Science, Masaryk University\\Kotlářská 267/2, Brno, 611 37, Czech Republic}

\maketitle

\begin{abstract}
Two interesting hypotheses about black holes have been proposed. The older one states that microscopic black holes can be accountable for the observed dark matter density. The newer one states that black holes are coupled to the expansion of the universe. Here, we combine those ideas and investigate the behaviour of expanding microscopic black holes. We observe two temperatures at which the radiation balances the expansion. While one of the balance points might be important in the analysis of primordial scenarios, the other would lead to a strong diffuse gamma radiation background, which is contradicted by the lack of observations. This establishes another indirect evidence disfavouring the hypothesis of cosmological coupling of black holes.
\end{abstract}

\keywords{Regular black holes; expanding black holes; Hawking radiation .}

\section{Introduction}

Two of the most astonishing findings of Einstein's theory of general relativity are the expansion of the universe and the existence of black holes. Each of the concepts is still being actively researched. In the case of black holes, a series of papers investigates models with a regular inner structure \cite{Frolov:1981mz, Poisson:1990eh, Hayward:2005gi, Ashtekar:2005cj, Bronnikov:2006fu, Chamseddine:2016ktu, BenAchour:2017ivq, Lan:2023cvz}, possibly effectively described as having a de Sitter core. 

Perhaps the most daunting open question related to the expansion of the universe is the nature of the force driving it today --- currently being described as dark energy, which is being studied thoroughly but is far from being properly understood \cite{Copeland:2006wr, Brax:2017idh, Peebles:2002gy}. 

Recently, several papers analyzed black holes and other relativistic objects coupled to the expansions of the universe \cite{Croker:2019mup, Croker:2019kje, Croker:2020plg}; in \refcite{Croker:2021duf}, the authors considered cosmological coupled black holes in the context of detection of gravitational waves. In \refcite{Croker:2019mup}, the Friedman equation was derived from the Einstein-Hilbert action with an explicit formulation of the averaging procedure. A startling consequence of the analysis was that relativistic compact objects, such as de Sitter-core black holes, should be expanding with the universe. An exciting scenario becomes possible if the size --- and therefore the mass --- of black holes scales with the cosmological scale factor. The energy of a black hole may grow as $\sim a^3$, which would cancel the $\sim a^{-3}$ decrease in the concentration of black holes in a proper volume. As a result, the energy contribution of black holes to the universe's energy budget would be constant and could be the driving force of its expansion. In \refcite{Croker:2019kje}, it was suggested that compact remnants of population III stars would correspond to the currently observed contribution of dark energy. Also, a recent observational study suggests the assumed scaling factor \cite{Farrah:2023opk} --- an intriguing picture began to emerge. 

So far, every black hole observed, directly or indirectly, had a mass above the limit set by the stellar collapse scenario. In the current stage of the universe, this seems to be the only way of creating a macroscopic black hole --- first being formed by a gravitational collapse of a stellar object and then growing either by accretion or merger with other black holes \cite{Mapelli:2020vfa, LIGOScientific:2016aoc, PortegiesZwart:1999nm}. However, it has been suggested that black holes of mass below the stellar limit could have been formed during the universe's earliest stages from local overdensities \cite{Carr:1975qj, Carr:2009jm, Carr:2020xqk}. While for stellar or supermassive black holes, the Hawking radiation is negligible, it would be crucial for sufficiently small black holes --- the absence of a signal of this type applies substantial restrictions on theories of small primordial black holes \cite{Barnacka_2012}. A hint of a subsolar black hole was reported recently in \refcite{Morras:2023jvb}.

The temperature of the Hawking radiation from a Schwarzschild black hole is inversely proportional to its mass. That means that as the black hole evaporates, it heats up, and the evaporation process accelerates. This would suggest that the presence of microscopic black holes can be disregarded, as suggested by Hawking \cite{Hawking:1974rv} for black holes with mass under $10^{15}g$. However, it is now understood that quantum effects can change this picture profoundly. For theories with a fundamental length scale $\lambda$, which is usually assumed to be the Planck length, the behaviour of microscopic black holes shifts drastically once their Schwarzschild radius reaches $r_H \approx \lambda$. In \refcite{Kovacik:2021qms}, it was shown that a large class of black holes that have a smeared singularity ––– so-called regular black holes --- have a vanishing Hawking temperature at this scale. This feature has been observed in particular examples before \cite{Dymnikova:2015yma, Nicolini_2005, Kovacik:2017tlg}. Instead of evaporating completely, they form a Planckian remnant, which is frozen and small enough not to interact with other fields; its cross-section is on the order of $\lambda^2$. Such objects were hypothesized to be plausible dark matter constituents \cite{Kovacik:2017tlg, MacGibbon:1987my, Chen:2002tu, Chen:2003bn, Bernal:2020kse, Green:2020jor}; recently, new details were discussed, either on the exact moment of their formation \cite{Kovacik:2021qms, Lehmann:2021ijf}, the scale of the underlying structure $\lambda$ \cite{DiGennaro:2021vev} or their M-theory or stringy aspects \cite{Borissova:2023kzq, Ho:2023tdq}.

This paper investigates how microscopic black holes would behave had they been expanding with the universe. We show that if small black holes exist and make a significant proportion of the dark matter, they would, due to the expansion of the space, produce a strong gamma signal, which is not being observed. We also discuss how the cosmological coupling of the black holes would affect primordial black hole formation.
\section{Small black holes}

Hawking deduced \cite{Hawking:1975vcx, Hawking:1974rv} that black holes have a temperature inversely proportional to their mass:

\begin{equation} \label{Hawking temperature}
 T = \frac{\hbar c^3}{8 \pi G k_{\scriptsize{B}} M}.
\end{equation}
Hawking radiation has a negligible effect on stellar black holes but may be substantial in late-age scenarios of the universe \cite{Aurell:2020ado}. However, small black holes have a temperature that tends to infinity in the final moments of the evaporation process. 

Primordial black holes (PBH) formed from matter overdensities in the universe's earliest stages and can have masses below the range of stellar black holes \cite{Carr:2020gox}; observing a gravitational wave signal from them might be within reach shortly either with space or ground-based detectors \cite{NANOGrav:2023gor, Branchesi:2023mws, LISA:2017pwj}. For such black holes, the radiation can be arbitrarily energetic --- lack of such signal places severe restrictions on the parameter space \cite{Barnacka_2012}. Recently, pulsar time array data were used to put constraints on the primordial power spectrum, which also limits the overproduction of PBH; however, additional data are required to strengthen these claims. Another possible probe of PBH is the analysis of dark-halo microlensing \cite{Delos:2023fpm}. 

So far, our discussion has only concerned singular black holes. Various theories considering the quantum effects predict that the singularity is blurred --- this is, for example, a rather general prediction of theories of quantum space. In \refcite{Kovacik:2021qms}, three different classes of spherically symmetrical regular matter densities were analysed:
\begin{eqnarray} \nonumber \label{densities}
\rho_1(r) &\sim & e^{-\left(r/\lambda\right)^q}, \\ 
\rho_2(r) &\sim & \left(1+\left(r/\lambda\right)\right)^{-q}, \\ \nonumber
\rho_3(r) &\sim & \left(1+\left(r/\lambda\right)^{q}\right)^{-1}.
\end{eqnarray}
Here, $r$ is the radial coordinate, $\lambda$ is the fundamental lengh-scale and $q$ is a model-dependent parameter, and $\lambda$ is a constant with the dimension of length that describes the scale of the quantumness of space. The previously considered cases of microscopic black holes \cite{Dymnikova:2015yma, Nicolini_2005, Kovacik:2017tlg} fall within $\rho_1$ with $q=1,2,3$. The similarity in their behaviour was extended in \refcite{Kovacik:2021qms} to all three classes with $q$ up to $15$. Therefore, the behaviour discussed in what follows can be considered robust.

\begin{figure}%
 \centering
\includegraphics[width=6cm]{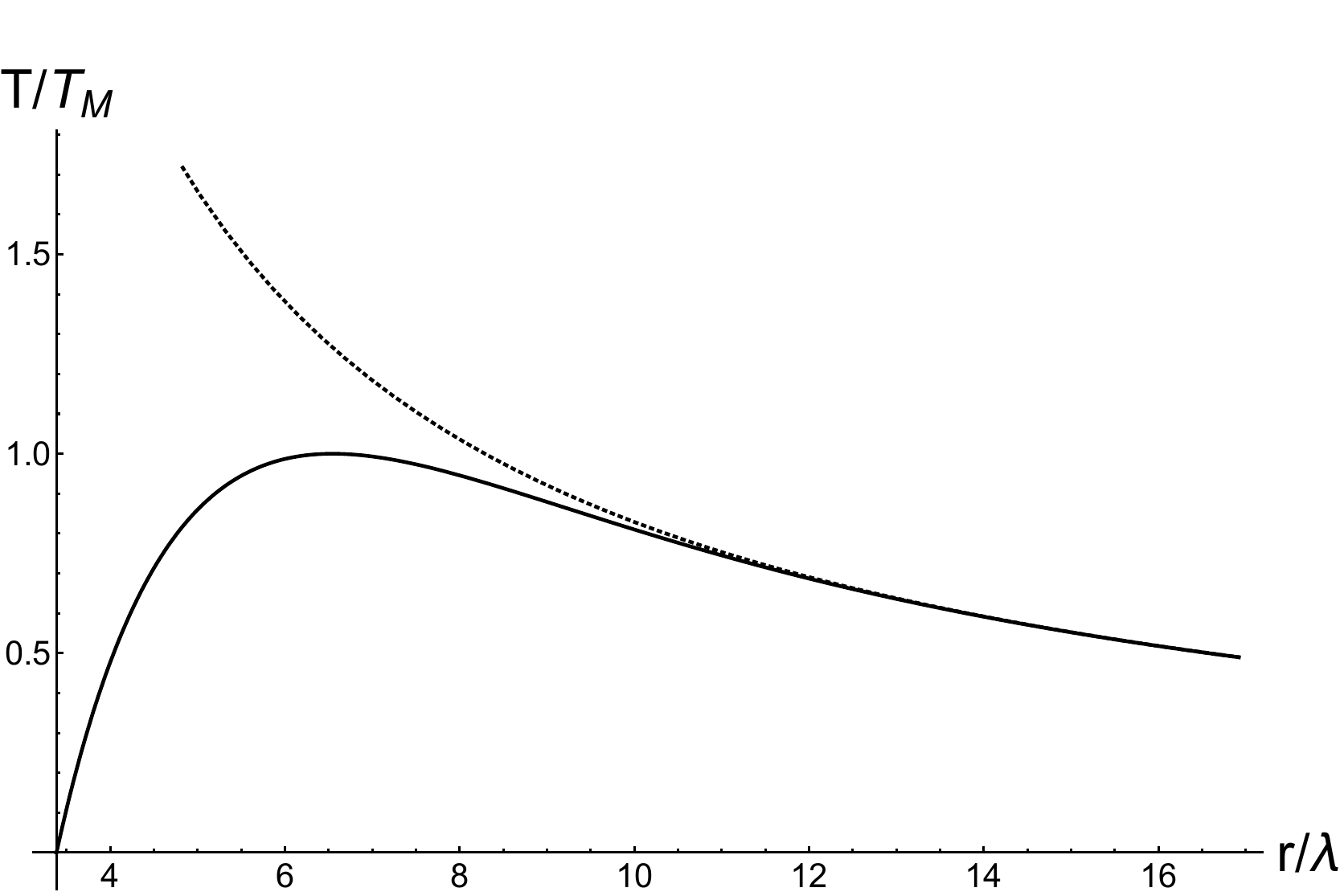} 
 \caption{\small{Comparison of the rescaled temperature profiles $T/T_M$ of Schwarzschild (dotted) and regular (solid) black holes. Values were obtained for the model where the mass density inside the black hole behaves as $\rho(r) \sim \exp\left(-r/\lambda\right)$ but the shape of the curve is general. In this case, the minimal horizon value is $r_{\scriptsize{0}} \approx 3.38 \lambda$ and $T_M \approx 0.01 \frac{\hbar c}{k_{\scriptsize{B}} \lambda}$.}}%
 \label{figure1}%
\end{figure}

It has been shown in this general setting that the temperature is regulated when the size of the horizon approaches the fundamental length scale, $\lambda$:
\begin{equation} \label{reduction of temperature}
 T_{\mbox{\scriptsize{reg}}} (m) = Q_{\scriptsize{\lambda}}(m) \times T_{\mbox{\scriptsize{sing}}} (m) , 
\end{equation}
where $T_{\mbox{\scriptsize{sing}}}$ is the temperature of a black hole with a singular mass density and  $T_{\mbox{\scriptsize{reg}}}$ is the temperature of a black hole with regular matter distribution. For the considered cases, $Q_{\scriptsize{\lambda}}(m)$ is a monotonic function that grows from $Q_{\scriptsize{\lambda}}(m_0)=0$ to $Q_{\scriptsize{\lambda}}(\infty) = 1$, see figure \ref{figure1}. Here, $m_0$ is the minimal mass of the black hole. For $m<m_0$, the object has no event horizons; for $m>m_0$, there are two, and the radiation temperature from the outer horizon is positive. For $m_0$, the two horizons overlap, and consequently, the temperature vanishes. Instead of evaporating with unbounded temperatures, black holes with blurred singularity have a maximal temperature below which they quickly freeze, and the radiation ceases. This happens as they reach the fundamental length scale size, which is assumed to be the Planck length --- so these objects are sometimes referred to as Planck size remnants or Planckian black holes \cite{Chen:2004ft, Ward:2006vw, Calmet:2014uaa}. They have a negligible cross-section and, therefore, do not interact with matter or radiation, so they do not accrete. Being massive and not interacting with ordinary matter or radiation makes them a viable dark-matter candidate. To make up for the observed dark matter density, their concentration needs to be of the order of $n \sim 10^{-20}\ m^{-3}$, see \refcite{Kovacik:2017tlg}, and unless the fundamental scale is far away from the Planck length \cite{DiGennaro:2021vev}; their production has to be completed at the age of the universe around $t\approx10^{-28}\ s$, see \refcite{Kovacik:2021qms}, \refcite{Lehmann:2021ijf}. 

\begin{figure}%
 \centering
\includegraphics[width=6cm]{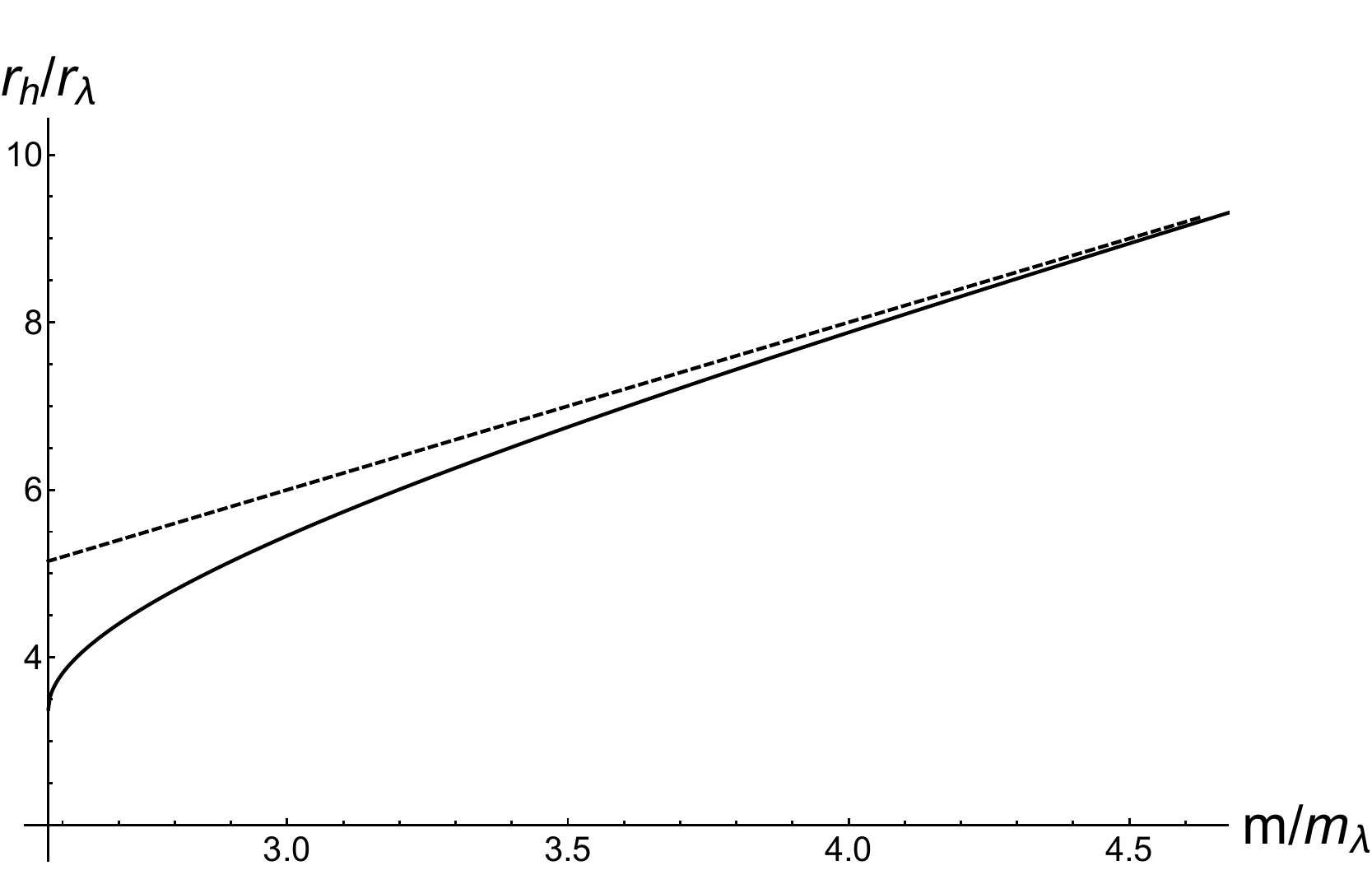} 
 \caption{\small{ Comparison of the event horizon radius of the Schwarzschild black hole, $r_h = 2 m$, (dotted) and a regular black hole (solid). Values were obtained for the model where the mass density inside the black hole is $\rho(r) \sim \exp\left(-r/\lambda\right)$. In this case, the minimal horizon value is $r_{\scriptsize{0}} \approx 3.38 \ \lambda$. The plot shows that taking $r_h = r_{\scriptsize{0}}$ for $m \approx m_{\scriptsize{0}}$ and $r_h = 2m$ elsewhere is justified --- the difference is a factor of the order of $1$.}}%
 \label{figure1b}%
\end{figure}

 There are two essential properties of regular black holes. The first is the form of the solution that resembles, for a nonrotating and electrically neutral source, the Schwarzschild solution for $r \gg \lambda$ while for $r \approx \lambda $, it tends to the de Sitter solution. This is the effect of a quantum space that prevents a further collapse and acts as an effective outward pressure. The second important property is the general shape of the temperature profile, as discussed in the previous paragraph.

 Therefore, we will assume this general behaviour and investigate the consequences of coupling a microscopic black hole to the universe's expansion. A similar case ignoring the radiation, therefore valid for large regular black holes, was investigated recently \cite{Cadoni:2023lum}. The focus of our analysis will be the interaction of two competing effects --- the expansion due to the expansion of space and their shrinking due to Hawking radiation. There is an implicitly time-dependent temperature where these two effects are balanced:
\begin{equation} \label{balancing}
 \left. \frac{dm }{dt}\right\vert_{\mbox{\scriptsize{exp}}} = 3 m  H = 4 \pi r_{\scriptsize{h}}^2 \sigma_B c^{-2} T^4 = - \left. \frac{dm}{dt}\right\vert_{\mbox{\scriptsize{rad}}}, 
\end{equation}
where we took the mass of the black hole to be coupled to the expansion of the universe as $m(t) = m(t_i)\left(a/a_i\right)^3$ as suggested in \refcite{Farrah:2023opk} and $r_{\scriptsize{h}}$ denotes the outer black hole horizon. Also, $H = \dot{a}/a$ is the time-dependent Hubble parameter.

Let us denote $T=T_{\scriptsize{B}}$ the temperature for which equation \ref{balancing} holds. A balance can be reached as long as $T_B < T_M$, where $T_M$ is the maximal temperature of a black hole. For the classes of solutions considered here, the maximal temperature is typically two or three orders below the Planck temperature; see \refcite{Kovacik:2021qms}. Because of the general form of the temperature profile $T(m)$ shown in figure \ref{figure1}, there are two possible masses at which the balance is reached; we denote them $m_{\scriptsize{1}}$ and $m_{\scriptsize{2}}$. Because the balance in equation \ref{balancing} is of the form $\alpha r_{\scriptsize{h}}^{-1} = T^4$, it holds that $T(m_{\scriptsize{1}})>T(m_{\scriptsize{2}})$. Typically $m_{\scriptsize{1}}$ is close to the minimal mass, $m_{\scriptsize{0}}$, where the temperature is zero but grows rapidly above it, and $m_{\scriptsize{2}}$ is considerably larger:
\begin{equation}
 m_{\scriptsize{0}} \lesssim m_{\scriptsize{1}} \ll m_{\scriptsize{2}}.
\end{equation}
We will now consider those two cases of balancing masses separately. 
\section{Balancing at $m_{\scriptsize{1}}$ and background radiation from expanding microscopic black holes}

As expressed in equation \ref{reduction of temperature}, the Hawking temperature vanishes as the minimal mass $m_{\scriptsize{0}}$ is reached. In cosmological models, the Planckian remnants with $m=m_{\scriptsize{0}}$ are assumed to be frozen and not radiating. However, if the black holes are coupled to the universe's expansion, a different balance point is reached; this is expressed in equation \ref{balancing}. Close to the minimal mass, $m \gtrsim m_{\scriptsize{0}}$, the temperature rapidly grows from zero. In \refcite{Kovacik:2017tlg}, it was calculated that the temperature close to this point behaves as
\begin{equation} \label{temperature(m)}
 T(m) \sim T_M \left(\frac{ m-m_{\scriptsize{0}}} {m_{\scriptsize{0}}} \right)^{1/2},
\end{equation}
where $T_M$ is the maximal temperature reachable by a black hole --- typically a few orders below the Planck temperature \cite{Kovacik:2021qms}. The numerical factor in this relation is model-dependent but typically is of the order of $1$; for example, for the regular matter density, $\rho_1(r)$ with $q=1$ as given in \ref{densities} is the numerical factor $4.09$. Also, there is a non-trivial factor in the relation between the mass and the horizon radius, $r_h$, which is of the same order and which we will ignore here, see \ref{figure2}. The balance is now reached at 
\begin{equation} \label{balancing2}
 3 \ m_{\scriptsize{1}} c^2 H = 4 \pi r_h^2 \sigma_B T_M^4 \left(\frac{m_{\scriptsize{1}}-m_{\scriptsize{0}}}{m_{\scriptsize{0}}}\right)^2.
\end{equation}

Since the mass dependence of Hawking radiation is different around the small mass $m_1$, see \ref{temperature(m)}, and the large mass $m_2$, where we can take $T(m) \sim m^{-1}$, so is different the nature of the balance. For the larger mass, the balance in equation \ref{balancing} is expressed in the form 
${\beta_{\scriptsize{1}}(m) = m - \alpha_1 m^{-2} = 0}$ which is unstable since ${\left. \frac{d \beta_{\scriptsize{1}}}{d m} \right\vert_{m_{\scriptsize{2}}} >0}$. An increase in mass leads to an increase in the growth rate and vice versa. However, the second balancing equation \ref{balancing2} is of the form $\beta_{\scriptsize{2}}(m) = m - \alpha_2 m^{2} = 0$ and since $\left. \frac{d \beta_{\scriptsize{2}}}{d m} \right\vert_{m_{\scriptsize{1}}} <0$ the opposite happens --- increasing mass leads to increased radiation, and the system is returned to its equilibrium in a Le Chatelier way. Equation \ref{balancing2} can be put in the form:
\begin{equation}
 \gamma\ m_{\scriptsize{1}} = \left(m_{\scriptsize{1}}-m_{\scriptsize{0}}\right)^2,
\end{equation}
where $\gamma = \frac{3 H m_{\scriptsize{0}}^2 c^2}{4 \pi r_{\scriptsize{0}}^2 \sigma_B T_M^4}$. For the current value of the Hubble parameter, we have $\gamma_{\scriptsize{0}} = 2.85 \times 10^{-61} kg$. The solution to this is, within the leading order approximation, $m_{\scriptsize{1}} = m_{\scriptsize{0}}+\sqrt{m_{\scriptsize{0}} \gamma _{\scriptsize{0}}}$, where the correction is indeed minimal $\frac{\delta m_1}{m_{\scriptsize{0}}} \approx 2.25 \times 10^{-27}$. If we evaluate the temperature as expressed in equation \ref{temperature(m)}, we obtain $T_{\scriptsize{B}} \approx 6.32 \times 10^{16} K$ --- which is considerably large.

That means that if the dark matter is made of Planckian remnants and those remnants are coupled to the universe's expansion, they should be radiating restlessly with the temperature of $6.32 \times 10^{16} K$. This translates into the energy of radiated particles $16$ orders below the Planck energy. If we compare the power at this temperature with the peak particle energy, we obtain that an emission happens with the frequency of $f \approx 0.013\ s^{-1}$.

In standard physical models, ultra-energetic are attenuated photons travelling through space due to interaction with low-energy background photons \cite{Amelino-Camelia:2016ohi, Protheroe:2000hp}. A distant, $z = 0.151$, the ultra-energetic signal has been observed recently, \cite{Amelino-Camelia:1997ieq, Brevik:2020cky, Li:2022wxc, Williams:2023sfk, Burns:2023oxn, Ripa:2023ssj, LHAASO:2023lkv}. Possible explanations of this phenomenon are particle physics beyond the standard model \cite{Troitsky:2022xso, Carenza:2022kjt} or non-trivial vacuum dispersion \cite{Zhu:2022usw, LHAASO:2024lub, Kovacik:2024ezt}. However, will focus only on local sources here; it was estimated in \refcite{Posti} that the mass of the dark matter in the inner part of the Milky Way is of the order of $\approx 10^{11} {M_\odot}$ which is $\approx 10^{49} m_{\scriptsize{\lambda}}$, recall that $m_{\scriptsize{\lambda}}$ is the Planck mass. If $\approx 10^{49}$ microscopic black holes were radiating thermal radiation with temperature $ 6.32 \times 10^{16}K$ with the peak energy $~ 15.3 TeV$, each having surface area $4 \pi \lambda^2$, they would be radiating $\approx 0.01$ high-energy particles isotropically every second, the total flux on Earth would be $1.82 \times 10^{5} m^{-2} s^{-1}$. Such an ultra-energetic gama background is, however, not being observed.

\section{Balancing at $m_{\scriptsize{2}}$ and primordial formation of black holes}
For sufficiently large masses, the regular and singular solutions become nearly indistinguishable. In the considered models is the minimal mass of the order of the Planck mass, $m_{\scriptsize{0}} \approx m_{\scriptsize{\lambda}} = \frac{\hbar}{\lambda c}$. At the larger balancing mass, we have  $m_{\scriptsize{2}} \gg m_{\scriptsize{0}}$, and so we can take the Hawking temperature as expressed in equation \ref{Hawking temperature}. The value of $m_{\scriptsize{2}}$ for which the balancing temperature is reached is given by the equation
\begin{equation}
 3 m_{\scriptsize{2}} c^2 H = \frac{\hbar c^6}{15360 \pi G^2 m_{\scriptsize{2}}^2 }.
\end{equation}
This can be expressed as
\begin{equation} \label{balancing mass}
 m_2 = \eta\ H^{-1/3},
\end{equation}
where $\eta = \left( \frac{c^{4/3} \hbar^{1/3}}{46080^{1/3}g^{2/3}} \right) \approx 109\ 807\ \mbox{kg}\ \mbox{s}^{-1/3}$. Today, with $H_0 \approx 2.2 \times 10^{-18}\ s^{-1}$, is the balancing mass $m_{\scriptsize{2,0}} \approx 8.41 \times 10^{10}\ kg$. 

This balancing point is unstable. If the mass increases, $m>m_{\scriptsize{2}}$, the expansion overtakes the radiation, the black hole keeps expanding, and the Hawking radiation becomes negligible. The black hole mass keeps scaling as $\sim a^3$, therefore mimicking the behaviour of the dark energy. In addition to that, the growth can be accelerated by the Bondi accretion \cite{Bondi:1944jm, Mack:2006gz, Ricotti:2007jk}. On the other hand, if the mass decreases, $m<m_{\scriptsize{2}}$, the evaporation takes over. It dominates until the point where quantum effects occur, and the temperature reduces according to equation \ref{reduction of temperature}. That means the black holes are driven from the balancing mass $m_{\scriptsize{2}}$. 

At the current stage of the universe, the observed black holes are well above the balancing mass \ref{balancing mass}. This discussion can be, however, relevant when considering the earliest stages of the universe, where some of the radiation and expansion scales were of the same order. The growth equation takes the form:
\begin{equation}
    \dot{m} = 3 H m - \Gamma m^{-2},
\end{equation}
and whether the radiation or expansion dominates depends on the Hubble parameter, $H$, see figure \ref{balancing}. As a result, the primordial black hole spectrum would be split into two parts, small black holes would be driven towards the mass $m_1$ while the large ones would grow indefinitely. Had we not concluded in the previous section that the existence of small microscopic black holes coupled to the expansion of the universe is contradicted by the lack of a strong gamma background, this would be a daunting model of primordial formation of both the dark matter (small black holes with mass $m_1$) and dark energy (large mass being driven away from $m_2$).

\begin{figure}%
 \centering
{\includegraphics[width=8cm]{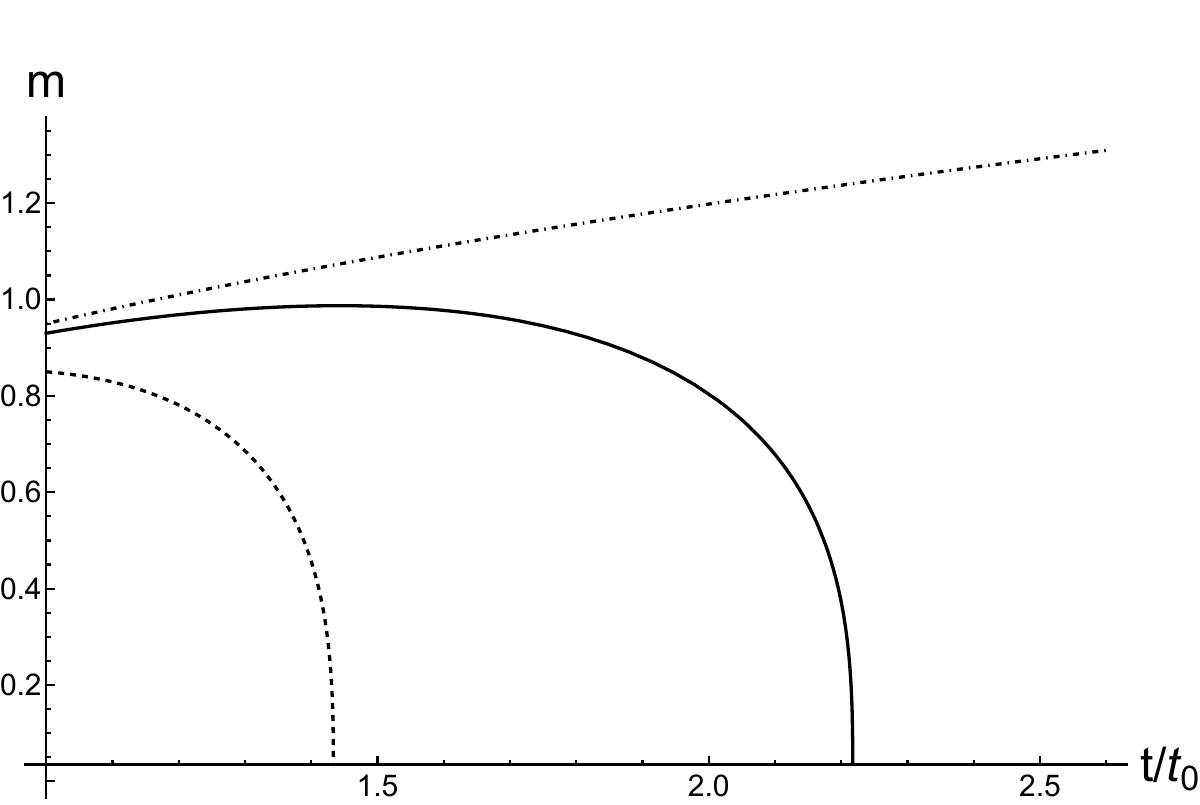} }
 \caption{\small{Mass evolution during a scenario when the expansion and radiation scales are comparable. There are three options: either the expansion dominates at every moment (dot-dashed line), it is surpassed at a certain point by the radiation (solid line), or the radiation always dominates (dashed line).}}%
 \label{figure3}%
\end{figure}

\section{Conclusion}
Microscopic black holes have been of some theoretical interest in recent years due to their peculiar properties --- the most important of which is their vanishing Hawking temperature at a sufficiently small scale. The black hole needs to have a regular, usually de Sitter, core for this modification to be present. This behaviour is a general feature of quantum space models; therefore, it is reasonable to assume that the relevant scale is that of Planck units. 

In this paper, we have discussed the interplay of two hypothetical concepts: black holes coupled to the expansion of the universe and microscopic black holes with regular matter distribution. On certain scales, the effects of expansion due to cosmological coupling and shrinking due to Hawking radiation are of the same magnitude. There are two different masses for which an exact balance is reached. They are not alike --- as balancing around the smaller mass is stable while around the larger one is unstable.

Expansion of the universe would be moving Planck-size black holes from the frozen state, making them radiate restlessly. If the dark matter at our galactic core was composed of radiating black holes, they would produce a gamma-background flux that has not been observed. We have considered only dark matter in our galaxy, but perhaps in this model, all dark matter in the universe would create a cosmic gamma background. Also, if the dark energy was sourced from PBH it would affect the delicate process of early nucleosynthesis and source primordial perturbations --- the exact observational signatures of this scenario require more research. Still, apriori, it doesn't seem easy to incorporate this into current models of the early-stage universe. 

There are a couple of possible explanations why this has not been observed: small black holes have never been formed, or black holes are not coupled to the universe's expansion and are therefore not forced to expand. However, there are other options; for example, Hawking radiation can be in the form of unobservable particles, or the fundamental scale that needs to be considered is not the Planck one. Also, details of the radiation process for microscopic black holes are not fully understood, and the scenario needs to be investigated using a proper model of quantum gravity. 

The other balancing point around the mass $m_2>m_1$ turned out to be unstable. The primordial mass spectrum would be split into two classes of black holes: one coupled to the expansion of the universe and the other of black holes that evaporated and their mass tended to $m_1$. 

In this paper, we have discussed the case of $m(t) = m(t_i)\left(a/a_i\right)^3$ as this is the option most frequently present in the literature. Other choices of the scaling coefficient would not change our results considerably. 

The bottom line is that abundant expanding microscopic black holes seem incompatible with current observations not detecting TeV photon background. This means that either Planck scale black holes do not exist or the black holes are not coupled to the universe's expansion, as other authors suggested recently for various reasons \cite{Parnovsky:2023wkc, Avelino:2023rac, Wang:2023aqe, Andrae:2023wge, Mistele:2023fds, Amendola:2023ays}. 

\section*{Acknowledgement}
This research was supported by VEGA 1/0025/23 and the MUNI Award for Science and Humanities, funded by the Grant Agency of Masaryk University. The author would like to thank J. Tekel and P. M\'esz\'aros for their valuable comments.

\bibliographystyle{ws-ijmpd}

\bibliography{References}

\end{document}